# Scalable terahertz generation by large-area optical rectification at 80 TW laser power

DOGEUN JANG,[1] CHUL KANG,[2,*] SEONG KU LEE,[2] JAE HEE SUNG,[2] CHUL-SIK KEE,[2] AND KI-YONG KIM[1,*]

[1] *Institute for Research in Electronics and Applied Physics, University of Maryland, College Park, MD 20742*
[2] *Advanced Photonics Research Institute, Gwangju Institute of Science and Technology, 123 Cheomdangwagi-ro, Oryong-dong, Buk-gu, Gwangju, 61005, Korea*
*Corresponding authors: iron74@gist.ac.kr; kykim@umd.edu*

**We demonstrate high-energy terahertz generation from a large-aperture (75 mm diameter) lithium niobate wafer by using a femtosecond laser with energy up to 2 J. This scheme utilizes optical rectification in a bulk lithium niobate crystal, where most terahertz energy is emitted from a thin layer of the rear surface. Despite its simple setup, this scheme can yield 0.19 mJ of terahertz energy with laser-to-terahertz conversion efficiencies of ~$10^{-4}$, about 3 times better than ZnTe when pumped at 800 nm. The experimental setup is up-scalable for multi-mJ terahertz generation with petawatt laser pumping.**

High-power terahertz (THz) generation is of great interest for its own study and potential applications in nonlinear THz spectroscopy and imaging, as well as exploration of terahertz driven extreme nonlinearities [1-4]. Optical rectification (OR) is widely used to produce single-cycle broadband THz radiation with femtosecond laser pulses. In general, OR is well suited for scalable THz generation—one can simply increase the OR crystal surface area with more input laser energy. Previously, a large-diameter (75 mm) ZnTe was used to produce 1.5 μJ of THz radiation with conversion efficiency of $3.1 \times 10^{-5}$ [5]. However, two-photon absorption in ZnTe with 800 nm pumping results in laser energy depletion and THz screening by charge carrier generation, fundamentally limiting maximum useable laser intensities on the crystal surface.

Organic crystals are also commonly used for high-energy THz generation due to their extremely high nonlinearities [6, 7]. However, organic crystals have low damage thresholds with large optical and THz absorption [6]. Generally, they are not favorable for optical pumping at 800 nm, and the growth of large-size organic crystals is also challenging. Nonetheless, monolithic arrays of partitioned organic crystals have been used to generate 0.9 mJ of THz radiation with 1,250 nm pumping [7].

Lithium niobate (LN) is another strong candidate for scalable THz generation because of its high nonlinearities and damage thresholds [8]. Owing to its large bandgap (~4 eV), LN is free from two-photon absorption at 800 nm. Moreover, high-quality, large-size LN wafers are commercially available with diameters beyond 152 mm. For efficient phase matching at 800 nm, a tilted pulse front method is routinely used for high-energy THz generation [9-14]. Recently, THz energy of 0.2 mJ was produced from a $68 \times 68 \times 64$ mm$^3$ sized LN prism with laser energy up to 70 mJ [14]. The prism-based geometry, however, induces non-uniform pump interaction due to its large tilt angle of 63° necessary for excitation at 800 nm. This results in poor THz beam quality. The condition can be even worse with large input beam sizes, which limits its potential use with multi-joule laser pumping.

In this paper, we examine scalable THz generation from LN in the form of a large-area wafer. Planar LN crystals are frequently used as OR materials, but their THz conversion efficiencies are not well known especially under extremely high laser energy and intensity conditions. Here we study laser chirp dependent terahertz generation and determine the ultimate THz conversion efficiency in LN wafers at 800 nm excitation.

A schematic of our experimental setup is shown in Fig. 1(a). In this experiment, a 150 TW Ti:sapphire laser capable of producing 25 fs, 4 J pulses at a repetition rate of 5 Hz is used. Due to possible cumulative degradation of the compressor gratings, the laser is operated in a single-shot mode. The laser pulse is p-polarized (horizontally polarized to the optical table) and clipped by an iris diaphragm to reduce the beam diameter to 50 mm with a maximum laser energy of 2 J after the iris. The laser spectrum is centered at 807 nm with a 60 nm bandwidth in full width at half maximum (FWHM) as shown in Fig. 1(b). The laser pulse duration is adjusted by adding group delay dispersion (GDD), either positive or negative, with an acousto-optic programmable dispersive filter (Dazzler, Fastlite) in the laser system. This provides a pulse width range of 25–900 fs (see Fig. 1(c)).

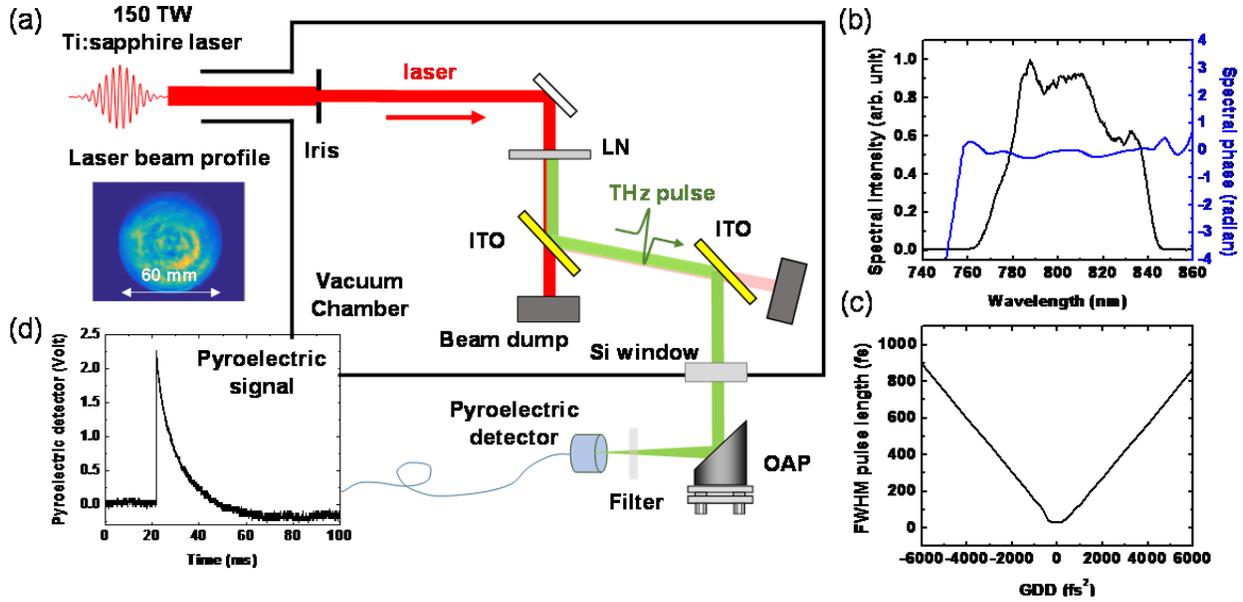

**Fig. 1.** (a) Experimental setup for high-energy THz generation from a large-aperture LN wafer. The inset shows the incident laser beam profile. (b) Laser spectral power (black line) and spectral phase (blue line). (c) Estimated laser pulse duration as a function of laser GDD. (d) Sample pyroelectric detector signal.

Three types of congruent LN are examined for THz generation—x-cut (0.5 mm and 1.0 mm thick) and y-cut (0.5 mm thick) wafers, all with 75 mm diameter. In the setup, the LN wafer is aligned with its extraordinary z-axis parallel to the laser p-polarization direction for maximal THz generation. This configuration yields p-polarized THz emission. To decouple the resulting THz radiation from its copropagating laser pulse, two optical windows coated with 250-nm thick tin-doped indium oxide (ITO) layers are used to reflect the THz radiation. To minimize reflection at 800 nm, the ITO-coated windows are tilted at Brewster's angle of 56°. This yields 800 nm and 1 THz reflection of 0.1% and 88%, respectively [15]. Double reflection with two parallel ITO windows provides overall THz reflection of 77%, with <0.0001% reflection at 800 nm. The reflected THz pulse is brought into air through a 10-mm thick, high-resistivity Si window that transmits ~50% of THz energy but completely blocks the attenuated laser pulse. The maximum laser fluence on the Si surface is 0.1 µJ/cm$^2$, sufficiently low enough to avoid potential charge carrier generation and subsequent THz screening.

The transmitted THz pulse is focused by a metallic, 90° off-axis parabolic (OAP) mirror with 102-mm diameter and 152-mm focal length. The THz energy is measured with a pyroelectric detector (Gentec, THz5D-MT-BNC) with filters including one or more calibrated THz attenuators (Tydex, ATS-5-50.8), a 1.0-mm thick Si filter, and a lowpass THz filter (Tydex, LPF-10.9-47) providing a cut-off frequency of 10 THz. The responsivity of the detector is measured at 800 nm with 0.9 µJ/V in a single-shot mode (see Fig. 1(d)) and converted to 1.8 µJ/V at <3 THz frequencies by using calibration data provided by Gentec.

First, LN crystal-cut and laser chirp dependence are studied. Figure 2 shows THz output energy as a function of laser GDD for the three types of LN. In general, at laser energy below 0.6 J, the shorter pulse duration yields the higher THz energy. However, at laser energy equal to or great than 1.2 J, the short pulse condition exhibits suppressed THz signals, with a bit longer pulse width conditions (100–200 fs) yielding more THz radiation. This is due to laser energy depletion via 3 photon absorption in LN occurring at high laser intensities. Note that three laser photons (1.5 eV each) can be simultaneously absorbed to exceed the bandgap of LN (~4 eV) and produce charge carriers. A similar behavior was previously observed with a tilted pulse front scheme using a thick LN prism [13, 14]. Also, for the same pulse duration, the negatively chirped ones, in general, yield slightly higher THz energy compared to the positive ones. This is due to dispersion in LN—a negatively chirped laser pulse undergoes temporal compression, producing higher laser intensity for THz generation, in contrast to a positively chirped one experiencing pulse stretching. Interestingly, the x-cut LN consistently yields more THz radiation than the y-cut under the same condition. This is unexpected because the laser is polarized along the z-axis in both cases, which in principle should yield the same linear and nonlinear responses.

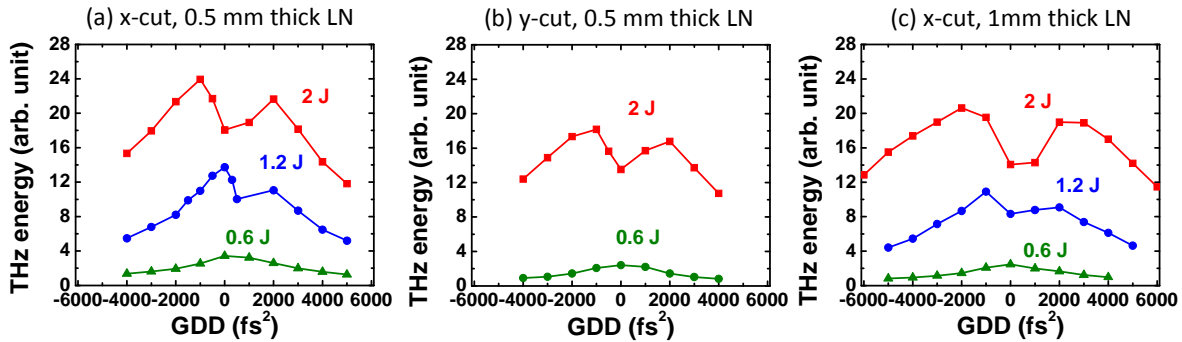

**Fig. 2.** THz output energy as a function of laser GDD with 0.6 J (green triangles), 1.2 J (blue circles), and 2 J (red squares) energy incident on a LN wafer of (a) x-cut and 0.5 mm thickness, (b) y-cut and 0.5 mm thickness, and (c) x-cut and 1 mm thickness.

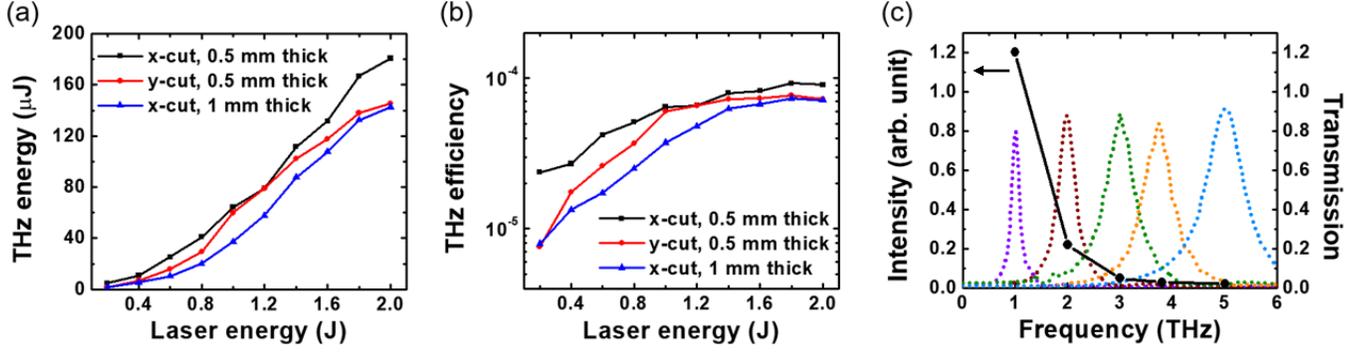

**Fig. 3.** (a) THz output energy and (b) laser-to-THz conversion efficiency as a function of laser energy. (c) THz spectral power (scatters) measured with THz bandpass filters (1, 2, 3, 3.8, and 5 THz) when produced from the x-cut, 0.5-mm thick LN at laser energy of 2 J with GDD of -1,000 fs$^2$. Co-plotted are the spectral transmission curves of the filters (dotted lines).

Figure 3(a) shows THz versus laser energy scaling for the three LNs. The y-axis represents THz energy estimated right after the LN crystal. Here the GDD value is fixed at -1,000 fs$^2$ for the 0.5-mm thick wafers and -2,000 fs$^2$ for the 1.0-mm thick LN. In general, the THz output energy continuously increases with input laser energy, also exhibiting saturation behaviors with high laser energy (>1.8 J). The corresponding laser-to-THz conversion efficiency is shown in Fig. 3(b). With the x-cut, 0.5-mm thick LN, the maximum THz energy produced is ~180 μJ with the conversion efficiency approaching 10$^{-4}$.

The radiation spectrum is characterized with 5 sets of metal-mesh bandpass filters (Thorlabs) [16] and shown in Fig. 3(c). All filters are calibrated, and their spectral transmission curves are plotted in Fig. 3(c). Here the THz spectrum (circles) is obtained by dividing the detected pyroelectric signal by the product of the width (FWHM) and height of each filter's transmission curve [16]. The resulting spectrum peaks at 1 THz and falls with increasing THz frequency. We also observed the presence of narrow band radiation at 15 THz radiation, expected from phase-matched optical rectification in LN [17]. Such radiation, however, is greatly attenuated by the 10-mm thick Si window and completely blocked by the longpass filter used in this experiment.

To simulate THz generation in a bulk LN, we solve 1-dimensional (1-D) coupled forward Maxwell equations (FME) given by

$$\frac{\partial E_T}{\partial \zeta} = -\left(\frac{\alpha}{2} + iD_T\right)E_T - i\frac{2d_{eff}\Omega}{c^2\varepsilon_0 n_T}\int_0^\infty \frac{I_L}{n_0}d\omega, \quad (1)$$

$$\frac{\partial E_L}{\partial \zeta} = -\left(\frac{\gamma}{2}I_L^2 + iD_L\right)E_L + i\frac{\omega n_2}{c}FT\{E_t I_t\}, \quad (2)$$

where $E_T = E_T(\Omega, \zeta)$ and $E_L = E_L(\omega, \zeta)$ are the THz and laser fields, respectively, $\zeta = z - v_g t$ is the coordinate moving at the laser group velocity $v_g$, $\alpha = \alpha(\Omega)$ and $D_T = k(\Omega) - \Omega/v_g$ are the absorption (by bulk material and charge carriers) and dispersion at THz frequency $\Omega$, $d_{eff}$ is the effective THz nonlinear coefficient, $n_0$ and $n_T$ are the refractive indices at laser and THz frequencies, and the term including $I_L = cn_0\varepsilon_0 E_L(\omega + \Omega, \zeta)E_L^*(\omega, \zeta)/2$ represents optical rectification by difference frequency mixing. The term including $\gamma$ and $I_L^2 = c^2 n_0^2 \varepsilon_0^2 |E_L(\omega, \zeta)|^4/4$ indicates laser depletion by 3 photon absorption, $D_L = k(\omega) - \omega/v_g$ is the dispersion at laser frequency $\omega$, $n_2$ is the nonlinear index of refraction, $E_t$ and $I_t$ are the laser field and intensity in time at $\zeta$, respectively, and $FT$ denotes Fourier transformation. The last term in Eq. (2) describes laser self-phase modulation (SPM). Initially, the laser field is set to be a Gaussian pulse centered at 800 nm with a FWHM bandwidth of 50 nm. This yields an initial pulse width of 19 fs when no GDD applied.

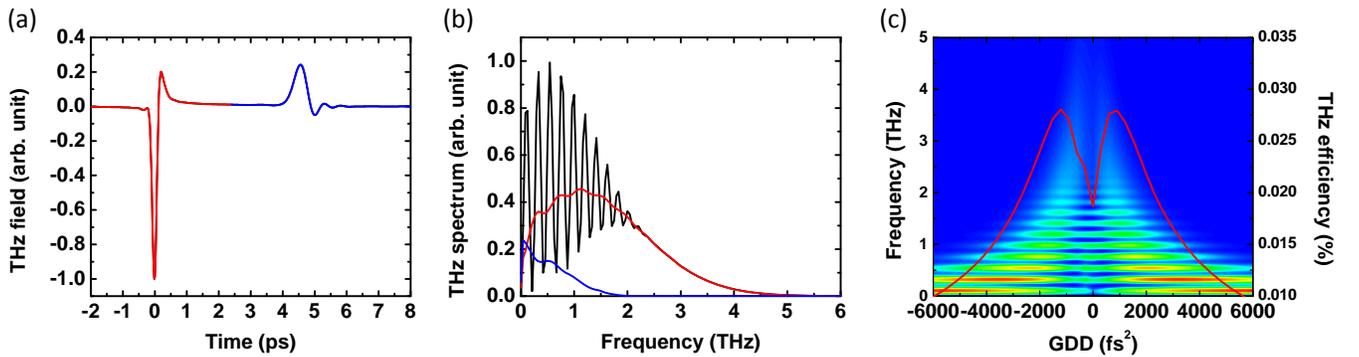

**Fig. 4.** (a) Simulated THz waveform produced from a 0.5-mm thick LN when irradiated by a laser pulse with 2 J and GDD of -1,200 fs$^2$. (b) Simulated THz spectral power for the leading (red), lagging (blue), and combined (black) THz pulses in (a). (c) Simulated THz waveform (false color) and laser-to-THz conversion efficiency (red line) as a function of GDD.

Figure 4(a) shows a simulated THz waveform captured 1 mm after exiting a 0.5-mm thick LN when irradiated by a laser pulse with an initial GDD value of -1,200 fs$^2$. It shows two emerging THz pulses separated by 4.6 ps. This is because phase matching is not satisfied throughout the entire thickness of LN. Instead, non-vanishing THz arises from mostly the front and rear surfaces of LN within one coherence length, $l_c = \lambda_{THz}/[2(n_{THz} - n_g)] = 50$ μm, from each surface. Here $\lambda_{THz}$ is the wavelength at 1 THz, $n_{THz} = 5$ is the refractive index at 1 THz, and $n_g = 2.2$ is the group index at 800 nm. The first pulse at $t = 0$ ps is produced from near the rear surface of LN, whereas the second pulse at 4.6 ps is created from near the front end.

Figure 4(b) shows the radiation spectrum (black line) obtained via Fourier transformation of the waveform in (a). The spectrum agrees well with our measurement in Fig. 4. The fast modulations observed at <2 THz are due to interference between two temporally separated THz pulses. The spectrum of the first pulse (red line) is much broader and stronger than that of the second one (blue line). This is because absorption $\alpha(\Omega)$ in LN increases with frequency $\Omega$, and the second pulse is affected more by this absorption. This also explains why the 0.5-mm thick LN produces more THz radiation than the 1.0-mm one, as shown in Fig. 1. With a thinner LN, the laser pulse experiences less energy depletion and SPM until reaching the rear surface, and the THz emitted from the front side undergoes more absorption.

Figure 4(c) shows simulated THz spectral power as a function of laser input GDD. It well reproduces the suppressed THz emission when excited by short pulses. Through the simulation, we confirm that this is mainly caused by laser energy depletion via 3 photon absorption, with little contribution from THz screening by charge carriers. The solid line represents the overall THz conversion efficiency with all spectral power integrated. It gives a maximum efficiency of 0.028%, not significantly different from our measured value (0.01%) given that many coefficients used in Eqs. (1-2) largely vary from literature to literature.

In conclusion, we have demonstrated scalable THz generation with Ti:sapphire laser energy up to 2 J using a 75-mm diameter LN wafer. This provides THz conversion efficiencies approaching $10^{-4}$ in a simple setup. The scheme is easily up-scalable by simply increasing the surface area of LN with more laser energy. For instance, a LN wafer with 152 mm diameter, currently commercially available, can accommodate laser energy of 18.6 J at laser fluence of 0.1 J/cm$^2$ and produce output THz energy up to 1.7 mJ. Potentially, contact grating [18], contact tilted reflection [19], or echelon step [20] structures can be fabricated on the front surface of a large LN to increase further the coherence length for phase matching and thus produce even higher THz energy.

**Funding.** Air Force Office of Scientific Research (AFOSR) (FA9550-16-1-0163); National Science Foundation (NSF) (1351455); This work was also supported by Gwangju Institute of Science and Technology (GIST) Research Institute (GRI) grant funded by the GIST in 2019 and the Institute for Basic Science (IBS) under IBS-R012-D1.

**Acknowledgment**. We thank Seung Woo Kang and Seong Chan Kang for their technical assistance.